\documentclass[prd,preprint,nofootinbib]{revtex4}
\usepackage{framed}
\usepackage{bbold}
\usepackage{slashed}
\usepackage{graphicx,hyperref}
\usepackage{amsmath}
\usepackage{amssymb}
\usepackage{epsfig}
\usepackage{hhline}
\usepackage{soul}
\usepackage{tabularx,pifont,multirow}
\usepackage{enumitem}
\usepackage{colordvi}
\usepackage{soul}
\usepackage{xcolor}
\usepackage{yfonts}

\newcommand{\nicktext}[1]{{\textcolor{black}{#1}}}\newcommand{\be}{\begin{equation}}
\newcommand{\ee}{\end{equation}}
\newcommand{\bear}{\begin{eqnarray}}
\newcommand{\eear}{\end{eqnarray}}

\newcommand{\MPl}{M_{\rm Pl}}

\definecolor{darkgreen}{rgb}{0,0.3,0.05}
\makeatletter                                                         %
\newcommand*\rel@kern[1]{\kern#1\dimexpr\macc@kerna}                  %
\newcommand*\widebar[1]{                                              %
  \begingroup                                                         %
  \def\mathaccent##1##2{                                              %
    \rel@kern{0.8}                                                    %
    \overline{\rel@kern{-0.8}\macc@nucleus\rel@kern{0.2}}             %
    \rel@kern{-0.2}                                                   %
  }                                                                   %
  \macc@depth\@ne                                                     %
  \let\math@bgroup\@empty \let\math@egroup\macc@set@skewchar          %
  \mathsurround\z@ \frozen@everymath{\mathgroup\macc@group\relax}     %
  \macc@set@skewchar\relax                                            %
  \let\mathaccentV\macc@nested@a                                      %
  \macc@nested@a\relax111{#1}                                         %
  \endgroup                                                           %
}                                                                     %
\makeatother                                                          %

\begin{document}

\rightline{KCL-PH-TH/2020-{29}, CERN-TH-2020-094,
~ACT-03-20, MITH-20-13}

\title{\vspace{-3cm}
\Large {\bf  Supercritical String Cosmology drains the Swampland}
~~\\}

\vspace{1cm}
\author{\bf John Ellis$^{a,b}$, Nick~E.~Mavromatos$^{a}$ and D.V.~Nanopoulos$^c$ \vspace{0.5cm}}

\affiliation{$^a$ Theoretical Particle Physics and Cosmology Group, Physics Department, King's College London, Strand, London WC2R 2LS, UK.
\vspace{0.25cm}\\
$^b$ Theoretical Physics Department, CERN, CH-1211 Gen\`eve 23, Switzerland; \\
National Institute of Chemical Physics \& Biophysics, R\"avala 10, 10143 Tallinn, Estonia
\vspace{0.25cm}\\
$^c$ George P. and Cynthia W. Mitchell Institute for Fundamental
 Physics and Astronomy, Texas A\&M University, College Station, TX
 77843, USA;\\
Astroparticle Physics Group, Houston Advanced Research Center (HARC),
 \\ Mitchell Campus, Woodlands, TX 77381, USA;\\ 
Academy of Athens, Division of Natural Sciences,
Athens 10679, Greece
}

\vspace{2cm}
\begin{abstract}
\vspace{0.05cm}
{\small The First and Second Swampland Conjectures (FSC \& SSC) are substantially modified in non-critical string cosmology,
 in which cosmic time is identified with the time-like 
Liouville mode of the supercritical string.  In this scenario the Friedmann equation receives additional contributions due to the non-criticality of the string.
These are potentially important when one seeks to apply the Bousso bound for the entropy of states that may become light as the dilaton takes on 
trans-Planckian values, as in a de Sitter phase, and restore consistency with the FSC and in at least some cases also the SSC.
The weak gravity conjecture (WGC) for scalar potentials
is saturated in the supercritical string scenarios discussed in this work, but
only if one uses the dilaton as appears in the string effective action, with a kinetic term that is not canonically normalised. 
In the case of a non-critical Starobinsky potential, the WGC is satisfied by both the canonically-normalised dilaton and the dilaton used in the string effective action.}

\end{abstract}
\maketitle

\section{Introduction \label{sec:intro}}

One of the outstanding challenges for string theory is to identify experimental measurements that
could test some characteristic prediction derived from the theory. Confronted with the tremendous 
gap between the primary formulation of string theory in multiple dimensions at a distance scale 
beyond the direct reach of observation, the conventional approach to this experimental challenge
has been to formulate and probe some effective field theory that describes four-dimensional
physics at accessible energy scales that is derived from (or at least motivated by)
string theory. Unfortunately, this approach has long been stymied by the enormous ambiguity in the 
choice of possible effective field theory. 

A promising way forward has been provided by the swampland conjectures~\cite{dSC,SC1,SC2a,SC2b,branden}, 
which postulate
some very general properties of the possible effective field theory, whose violation would lead
to the conclusion that (our current understanding of) string theory does not describe Nature.
This line of attack is particularly intriguing because some aspects of the swampland conjectures
appear to be in tension with the appearance of a cosmological constant in present-day
cosmology as well as the commonly-studied models of inflation in the early Universe,
and may raise questions about the formulation of spontaneous symmetry breaking in the
Standard Model of particle physics~\cite{higgs}.

One possible way to resolve these issues may be to abandon the conventional effective field
theory description of physics in the infra-red limit at energies far below the putative string
scale. Effective field theory emerges naturally from critical string theory via its formulation
in terms of theory on the string world sheet. We have long argued that
conformal field theory is just one aspect of the underlying string theory, and that critical string
theory should be regarded as embedded in a broader framework that includes supercritical
string scenarios in which there is a time-like Liouville mode that can be identified with cosmic
time in scenarios for an evolving Universe~\cite{emn,nc,diamant,timeRG,ncaccel}.

As we have discussed previously, the non-trivial dynamics of such a Liouville mode modifies
the equations of motion of a conventional effective field theory. In particular, as we emphasise
in this paper, in the cosmological context the supercritical dynamics introduces extra terms
into the Friedmann equation describing the evolution of the Universe, reopening the possibility
of a stringy description of inflation in the past and/or dark energy today.

We recall in this context that an essential feature of perturbative critical string theory is the
existence of a perturbative S-matrix. This requirement appears to exclude a de Sitter background, 
since it does not accommodate asymptotic states because of the presence of a horizon~\cite{dsSmatrix}.
On the other hand~\cite{nc,ncaccel,dollarS}, supercritical strings are formulated without requiring the
existence of an S-matrix~\cite{dollarS}, and hence can accommodate both an early near-de Sitter 
epoch from which the Universe exits smoothly and another  near-de Sitter epoch at late
times.~\footnote{In some supercritical string cosmologies~\cite{ncaccel},
the Universe approaches at large cosmic times a linearly-expanding universe~\cite{aben}
that is characterised by a logarithmic dependence of the dilaton on cosmic time, in which case
the perturbative S-matrix can be well-defined asymptotically~\cite{dollarS}.}

The outline of this paper is as follows. In Section~\ref{sec:swamp} we outline relevant aspects
of the swampland conjectures (SC). Then, in Section~\ref{sec:rvm} we examine the swampland
from the point of view of non-critical string theory. Section~\ref{sec:WGC} provides a parallel
discussion of the weak gravity conjecture (WGC) in the framework of non-critical string theory.

\section{The Swampland Conjectures \label{sec:swamp}}

There are several Swampland Conjectures (SC) that bear upon  
the possible embedding of a de Sitter (dS) solution in a low-energy effective field theory 
arising from a microscopic string theory model that is supposed to provide a self consistent  UV completion.\\

$\bullet$ \underline{The Distance in Field Space Conjecture (DFSC)}~\cite{dSC}

When considering a scalar field $\phi$ within the effective low-energy field theory (EFT) of some string theory, 
which is displaced by a `distance'  $\Delta \phi$ in  field space from some initial value, the validity of the EFT is guaranteed provided
\be\label{dSC1}
\kappa \, |\Delta \phi | \lesssim c_1 > 0 \, , 
\ee
where $c_1 $ is a positive constant of order ${\mathcal O}(1)$, and $\kappa=\frac{1}{\MPl}$ is the gravitational constant in four space-time dimensions, where $\MPl = 2.4 \times 10^{18}$~GeV  is the reduced Planck mass.
In general, when \eqref{dSC1} is violated towers of string states descend from the UV, becoming light and thus `contaminating' the EFT,  jeopardising any physical conclusions that may be derived from it. \\

$\bullet$ \underline{The First Swampland Conjecture (FSC)}~\cite{SC1}

This conjecture restricts the self-interactions of scalar field(s) in an EFT stemming from string theory via the gradient  of the scalar potential 
$V$ in field space, so that the EFT is valid:
\be\label{SC1a}
 \frac{|\nabla V|}{V} \gtrsim c_2 \, \kappa > 0 \, ,
\ee
where $c_2$ is a positive (dimensionless) constant of $\mathcal O(1)$.  The gradient in field space is in the multicomponent 
space of scalar fields $\phi_i$, $i=1,\dots N$ that the EFT contains. 

The constraint \eqref{SC1a} can be derived from the DFSC~ \eqref{dSC1}, using 
entropy in de Sitter  space~\cite{SC2a},
by using the Bousso bound~\cite{bouso}, which is in this case saturated by the Gibbons-Hawking entropy~\cite{GH}. 
Specifically, let us consider an EFT of a quintessence field $\phi(t)$ in which, say, the latter increases with
cosmic time, i.e., $\phi (t)$ has $\dot \phi > 0$~\footnote{This is a convention for the definition of the flow, which is made for concreteness, see the explicit examples below.}.
As the time evolves, if the field lies in regions in which \eqref{SC1a} is violated, 
towers of string states with masses $m \sim \exp (-a |\Delta \phi|)$, where $a > 0$ is a constant with mass dimension $-1$ that depends on the details of the microscopic string theory,
become lighter and lighter and descend from the UV-complete theory to contaminate the EFT as the `distance' $\Delta \phi$ from the initial point increases. 

The number of these effective light degrees of freedom $N(\phi)$ depends on the value of the field $\phi$ at any moment in time, and can be parametrised as~\cite{SC1}:
\be\label{tower}
N(\phi) = n(\phi)\, \exp (b \, \kappa \, \phi) \, ,
\ee
where $n(\phi)$ indicates the number of states in the tower that are becoming light, and $b$, like $a$, is another positive constant that depends on the mass gap and other details of the underlying string theory. Since, according to the distance conjecture, more states become light as $\phi$ grows, one has to take the (positive) function $n(\phi) > 0$ to be monotonically increasing with $\phi$, i.e. $\frac{d n(\phi)}{d \phi} > 0$.
 
In an accelerating  Universe, characterised by a Hubble parameter $H$, which for (near-)de Sitter-times is (approximately) constant,  
the entropy of this tower of string states increases with the Hubble horizon $1/H$ as
\be\label{entropy}
S_{\rm string~states} (N, H^{-1}) = N^\gamma \, (\kappa H)^{-\delta} \, ,
\ee
where $\gamma, \delta > 0$. On the basis of  string  examples, the authors of \cite{SC2a} argued that $0 < \delta \le 2$.   
If the tower of string states that are becoming light behave effectively as point particles, then $\delta=0$, 
but this is not necessarily the case, and the value of $\delta$ depends on the underlying microscopic string theory.

Since the expanding Universe is characterised by the presence of the Hubble horizon of area $A=4\pi H^{-2}$, 
the entropy is bounded by Bousso's covariant entropy constraint applied to a cosmological background~\cite{bouso}: $S_{\rm string~states} (N, H^{-1}) \le \frac{1}{4} A$, 
which is nothing other than the Gibbons-Hawking entropy~\cite{GH}:
\be\label{GH}
N(\phi)^\gamma \, (\kappa H)^{-\delta}  \le 8\pi^2 (\kappa H)^{-2} \, ,
\ee
from which we derive
\be\label{GH2}
H^2 \le \frac{1}{\kappa^2} \Big(\frac{8\pi^2}{N(\phi)^\gamma}\Big)^{2/(2-\delta)} \, .
\ee
If the potential energy of the scalar field dominates the energy density of the Universe, 
and one assumes the standard Friedmann equation
\be\label{friedmann}
H^2 \simeq \frac{\kappa^2}{3} V \, ,
\ee
i.e., the kinetic energy 
of the scalar field responsible for inflation is ignored, compared to the potential energy, then 
we can use \eqref{friedmann} to rewrite \eqref{GH2} as 
\be\label{friedmann2}
\frac{\kappa^4 \, V}{3} \le \Big(\frac{8\pi^2}{N(\phi)^\gamma}\Big)^{2/(2-\delta)} \, .
\ee
After some straightforward manipulations, and taking into account the fact that the derivative of the 
potential with respect to the field $\phi$ is negative~\footnote{Here we adopt the conventions of, e.g., the work in \cite{ncstaro},
as we are interested in the flow of the field from large values, where the de Sitter phase occurs, to small values, where the exit from inflation takes place.}, we find that 
\be\label{eSC1b}
\frac{|V^\prime |}{V} \ge  \frac{2}{2-\delta} (\ln [N^\gamma])^\prime \, ,
\ee 
where the prime denotes the derivative with respect to the field $\phi$. Using \eqref{tower}, one then obtains
\be\label{eSC1a}
\frac{|V^\prime|}{V} \, \ge \, \frac{2\, \gamma}{2-\delta} \, \Big(\frac{n^\prime}{n}
 + b \, \kappa \Big) \, > \, \frac{2\, b\, \gamma}{2-\delta} \, \kappa \, ,
\ee 
where the last inequality follows from the fact that $n^\prime > 0$. Comparing \eqref{eSC1b} with \eqref{SC1a}, we obtain the 
FSC with $c_2 = \frac{2\, b\, \gamma}{2-\delta} > 0$, provided that the parameters $b\, ,\gamma $ and $\delta$ 
are such that  $c_2$ is of order $\mathcal O(1)$.~\footnote{As already remarked, $\delta =0$ if only point-like states are included in the tower of states $N(\phi)$.}

The FSC \eqref{SC1a} rules out slow-roll inflation, since it implies that the parameter $\epsilon = (\sqrt{2}\, \kappa)^{-1} \,  (|V^\prime|/V)$ is of order one, 
in conflict with phenomenologically successful inflationary models~\cite{planck}.

In fact, it was argued in \cite{SC2a,SC2b} that the FSC does not necessarily hold, but that consistency of the EFT requires 
{\it either} the constraint \eqref{SC1a}, {\it or} \, :\\

$\bullet$ \underline{The Second Swampland Conjecture (SSC)}~\cite{SC2a,SC2b}

According to the SSC, the minimum eigenvalue of the Hessian in theory space ${\rm min}(\nabla_i \, \nabla_j V)$, 
with the notation $\nabla_i$ denoting the gradient of the potential $V$ with respect to the scalar field $\phi_i$, 
should satisfy the following constraint near the local maximum of the potential $V$, provided there is one:
\be\label{SC2}
\frac{{\rm min}(\nabla_i \, \nabla_j V)}{V} \, \le \, -  c_3 \, \kappa^2 \, < 0 \, ,
\ee 
where $c_3 > 0$ is an appropriate positive (dimensionless) constant that is $\mathcal O (1)$. 

We note that, for single-field inflationary models, the condition \eqref{SC2} would 
also be incompatible with the smallness in magnitude of the second of the slow-roll parameters $\eta$, as required  for conventional slow-roll inflationary models. 

We also remark that in case of models for which the SSC applies but {\it not} the FSC, the entropy-bound-based derivation 
of FSC still holds, but for a range of fields further away from the regime for which the local maximum of the potential occurs
(as required for the implementation of the SSC). The {\it critical value (range)} of the field magnitude for the entropy-bound 
implementation of the FSC depends on the underlying microscopic model, upon which 
the parameters $\gamma, b$ in the bound \eqref{eSC1a} also depend.  

\section{Supercritical String Theory and the Swampland \label{sec:rvm}}

 The central point of our discussion is the modification of the First Swampland Conjecture (FSC) \eqref{SC1a} in supercritical string theory, 
 in which the right-hand-side (r.h.s.) of the Friedmann equation of motion \eqref{friedmann} receives additional contributions due to 
 non-critical string degrees of freedom~\cite{nc,diamant}, which arise from identifying time as the Liouville mode of the supercritical string~\cite{timeRG}.  
 To see this, we consider a concrete but representative example of a supercritical string cosmology where the Liouville mode 
 is identified with cosmic time, and the dilaton plays the role of a quintessence field. 
 The scenario is discussed in detail in~\cite{diamant}, and here we simply outline the main results relevant for our purposes.
 
 \subsection{The Supercritical String Cosmology Approach} 
 
Before going into calculational details, it is important to recall briefly the underlying philosophy of the non-critical approach to
string cosmology of~\cite{nc}. This {\it necessarily} involves infinite towers of stringy states that 
fail to decouple from the light degrees of freedom. In the approach of \cite{nc} these string states provide
a non-trivial \emph{environment} for the low-energy effective field theory
(EFT), whose description incorporates only local field theory modes. 
When the string is {non-critical}, the world-sheet Weyl-anomaly coefficients, which appear as the renormalization-group 
$\beta$-functions of the couplings/target-space fields $g^i$ of the $\sigma$-model that describes string propagation, 
are no longer zero as in a conformal field theory corresponding to a critical string model.
Instead, they are determined by a standard Liouville-dressing procedure:~\footnote{For notation, conventions and more details 
we refer the reader to~\cite{nc} and references therein.}
\be
g^{i\prime\prime} + Q \, g^{\prime} = - \beta^i (g)  \ne 0 \, ,
\label{liouville}
 \ee
 where the $g^i$ are Liouville-dressed couplings/background fields in the $\sigma$-model, 
 which has $d+1$ coordinates, of which $d$ are space-like coordinates in target space, and one is time-like and interpreted as the target time. 
 The prime in \eqref{liouville} denotes differentiation with respect to the world-sheet zero-mode $\rho$ of the Liouville field 
 $\rho (\sigma)$, where $\sigma$ denotes collectively the dependence on world-sheet coordinates, which  is identified
with the target time in the approach of~\cite{nc}. The Liouville mode itself is viewed as a dynamical renormalization-group scale, which is 
local on the world-sheet and promoted to a fully-fledged quantum field on the two-dimensional world-sheet geometry. 
The right-hand side of \eqref{liouville} {\it vanishes}  in critical string theories, and this absence 
provides the standard target-space equations of motion of local fields in a low-energy string EFT.  
The identification with the target time of this extra coordinate in the target space of the string, 
which in supercritical strings has a time-like signature,  imposes a constraint on the $(d+2)$-dimensional target manifold, 
such that there is a single physical time.

A prototype that illustrates how towers of string states may affect the Weyl-invariance conditions and leading to Liouville dressing 
is the cigar-type black-hole background in two-dimensional string theory~\cite{witten}, which may be embedded
in higher dimensions~\cite{emnBH}. As observed in \cite{emn}, such a stringy background leads to the mixing with light modes
of infinite towers of topological modes of (non-propagating) {\it massive string states} with discrete momenta.
If they are not taken into account, the truncated $\sigma$-model that includes only backgrounds from the massless string states 
(the graviton and dilaton in this example) is no longer {conformally invariant}, rendering the string non-critical.
Liouville dressing is then required to restore world-sheet conformal invariance, and thus independence of the 
target-space physics from world-sheet dynamics.~\footnote{Studies of such stringy black holes, and the mixing of the environment of 
higher string modes, prompted the authors of \cite{emn} to discuss quantum coherence issues and modifications 
of naive quantum mechanics within such an EFT framework, identifying the Liouville mode with time
and mapping the problem to a suitable supercritical string.} 

Other examples of such mixing between massive and massless states occur in D-brane theories~\cite{emndbranes,ncstaro}, when one 
takes into account the space-time distortions induced by the recoil of such extended objects during scattering with string states
representing low-energy matter. Such an approach can also lead to a time-dependent vacuum energy, 
with a dark energy-like equation of state that exhibits temporal dependence (relaxation)~\cite{emnLt}~\footnote{For phenomenological consequences of relaxation models for the cosmological dark energy in the context of non-critical string cosmologies of \cite{aben}, see \cite{ln}. We also note that super-critical string cosmologies of the type we focus our attention upon here~\cite{diamant} could also contribute to the alleviation of the apparent $H_0$ tension~\cite{H0tension}.}. This
cosmological situation is similar in spirit to the effective field theory model of a running vacuum~\cite{rvm}, 
but involves brany objects whose recoil induces the temporal dependence via non-criticality of the underlying string theory.~\footnote{An 
embedding of the running vacuum model in {critical string theory} and its cosmological implications have been discussed in \cite{bmas}.}. 

When applied to specific non-critical string models~\cite{nc,diamant,ncaccel}, the Eqs.~\eqref{liouville} 
result in a coupled system of equations that constitute an extension of the cosmological equations of
standard critical string cosmologies. The simplest and most representative systems we deal with here, which suffice for our purposes, 
contain dilaton and metric backgrounds only. In more realistic string models~\cite{ncaccel} one considers
other fields, such as fluxes, string moduli, antisymmetric tensors, {etc.}, which provide additional equations of  motion and constraints among the relevant parameters of the models, but do not affect our basic conclusions on the fate of the First Swampland Conjecture \eqref{SC1a} in the non-critical cosmology framework.

We conclude this summary by commenting on the relation between the approach adopted here
and the linear-dilaton string cosmology of \cite{aben}. The latter is regarded as a critical, ``equilibrium'' string configuration,
which is described by a conformal field theory on the world-sheet, even though the internal central charge is non-critical. 
In the approach of \cite{nc}, the presence of the Liouville mode and its identification with target time provides a 
{non-equilibrium} ``transient'' period for the associated string universe, which may in some instances tend 
asymptotically~\cite{diamant,ncaccel}  to the equilibrium critical string model of \cite{aben}.

\subsection{A Representative Supercritical String Cosmology Model and the Swampland Conjectures} 

Considering metric and dilaton cosmological backgrounds, along with some generic matter and radiation (denoted by the suffix $m$), one finds  
the following cosmological equations in the Einstein frame, upon identifying the (supercritical) Liouville mode 
with the target time:~\footnote{For notational convenience in the rest of the article, unless otherwise stated, we work in units with $\kappa=1$.}
 \bear
&&3 \; H^2 - {\tilde{\varrho}}_m - \varrho_{\phi}\;=\; \frac{e^{-\sqrt{2} \phi}}{2} \; \tilde{\cal{G}}_{\phi} \, , \nonumber  \\
&&2\;\dot{H}+{\tilde{\varrho}}_m + \varrho_{\phi}+
{\tilde{p}}_m +p_{\phi}\;=\; \frac{\tilde{\cal{G}}_{ii}}{a^2} \, , \nonumber  \\
&& \ddot{\phi}+3 H \dot{\phi}+ \; \frac{\partial {{V}}_{all}(\phi) }{\partial \phi}
- \frac{1}{\sqrt{2}} \;( {\tilde{\varrho}}_m - 3 {\tilde{p}}_m )= 
+ \frac{3}{2}\; \frac{ \;\tilde{\cal{G}}_{ii}}{ \;a^2} +  \,
\frac{e^{-\sqrt{2} \phi}}{2} \; \tilde{\cal{G}}_{\phi}  \; ,
\label{eqall}
\eear
where $\tilde\varrho_m$ and $\tilde p_m$ are the conventional matter density and pressure,
$\varrho_{\phi}$ and $p_{\phi}$ are the dilaton energy density and pressure, and:~\footnote{We note for completeness~\cite{diamant} that the function
${\tilde{\cal{G}}}_{00}$, which is the $00$ component of
${\tilde{\cal{G}}}_{\mu \nu}$,  vanishes because the corresponding
component of the metric is constant.}
\bear \label{Gfidef}
\hspace{-1.5cm}
&&\hspace{-0.9cm}
 \tilde{\cal{G}}_{\phi} \;= - \; e^{\;\sqrt{2} \phi}\;\Big(  \frac{1}{\sqrt{2}}\, \ddot{\phi} + \frac{1}{2} \, 
{\dot{\phi}}^2 + \frac{Q(t)}{\sqrt{2}}\, e^{-\phi/\sqrt{2}} \dot{\phi}\Big) \, , \nonumber \\ && \hspace{-0.9cm}
\tilde{\cal{G}}_{ii} \;=\; 2 \;a^2 \;\Big(\; -\frac{1}{\sqrt{2}}\, \ddot{\phi} - \frac{3}{\sqrt{2}} H \, \dot{\phi}
+ \frac{1}{2} \, {\dot{\phi}}^2 + ( 1 - q ) H^2 -\frac{1}{\sqrt{2}}\, Q(t)\, e^{-\phi/\sqrt{2}} ( \dot{\phi}-\sqrt{2} H )\;\Big) \, ,
\eear
where $Q^2(t)$ is the central-charge surplus, 
$a$ the cosmic scale factor and $q$ the cosmic deceleration, 
and we note that only two of the equations \eqref{eqall} are independent. 
We have used in the above a dilaton field with a canonically-normalised kinetic term, following the notation in \cite{ncstaro}. In particular,
the dilaton $\phi$ used in \cite{ncstaro} and appearing in \eqref{eqall}  is related to the standard dilaton $\Phi$ used in \cite{aben} and \cite{nc,diamant} by 
\be\label{standard}
\phi=-\sqrt{2}\, \Phi \, .
\ee
In our conventions, the slope of the potential is positive, as the potential increases for increasing $\phi$, 
until it assumes an approximately constant value for large values of $\phi$. 

In the convention \eqref{standard}, the string coupling is given by 
\be\label{scoupl}
g_s = g_s^{(0)} \, e^\Phi = g_s^{(0)} \, e^{-\phi/\sqrt{2}} \, , 
\ee
and, as we discuss below, in all our non-critical cosmology examples~\cite{aben,nc,diamant,ncstaro}, the string coupling is {\it weak}, 
including during the inflationary era, in the sense that the dilaton takes on large (trans-Planckian) positive values.
Indeed it is an increasing function of the cosmic time. One can take $g_s^{(0)} < 1$ to guarantee string-loop perturbation theory 
in all the phenomenologically relevant epochs of the Universe evolution, from early cosmology to the current era.

The dots above a quantity in \eqref{eqall} denote derivatives with respect to the Einstein-frame cosmic time, which is the standard time used in cosmology to compare models with observations~\cite{planck}. 
The r.h.s. of these equations are provided by the terms that are associated with  the non-critical string   
behaviour. The time derivatives appearing on the r.h.s. are essentially derivatives with 
respect to the Liouville mode, which is the zero mode of the world-sheet renormalisation scale 
and has been {\it constrained} to be identical to the cosmic time~\cite{nc}.
Such  terms are absent in critical string cosmologies, including the asymptotic situation met in \cite{aben}.

We ignore for our purposes here the matter/radiation contributions denoted by the suffix $m$, 
as we are only interested in early (inflationary) epochs, in particular the validity of the Swampland Conjectures 
for de Sitter space time and the EFT in the context of non-critical string cosmologies, and hence set
$\tilde\varrho_m=\tilde p_m=0$.

The reader should notice the {\it dissipative} terms proportional 
to $Q(t) \, \dot{\phi}$ in~(\ref{Gfidef}), where $Q(t)$ is the central charge surplus, that depends in general on the cosmic time~\cite{diamant}. 
There are many explicit ways to generate such as a surplus,
and we shall come back to explicit examples discussed below. 
In general, the variation of the
central charge deficit 
$Q(t)$ with the cosmic time is provided~\cite{diamant,ncaccel} by the Curci - Paffuti
equation~\cite{curci}, which expresses the renormalisability 
of the world-sheet theory. To leading order in an $\alpha '$ expansion,
to which we restrict ourselves in this work, 
this equation is given by
\bear \frac{d \tilde{\cal{G}}_{\phi} }{d t} \;=\; - 6\; e^{\;\sqrt{2}
\phi}\;( H - \frac{1}{\sqrt{2}} \, \dot{\phi} ) \; \frac{ \;\tilde{\cal{G}}_{ii}}{ \;a^2}
 \label{CXP} \eear
in the Einstein frame.

\begin{figure}[ht]
       \includegraphics[width=0.8\textwidth]{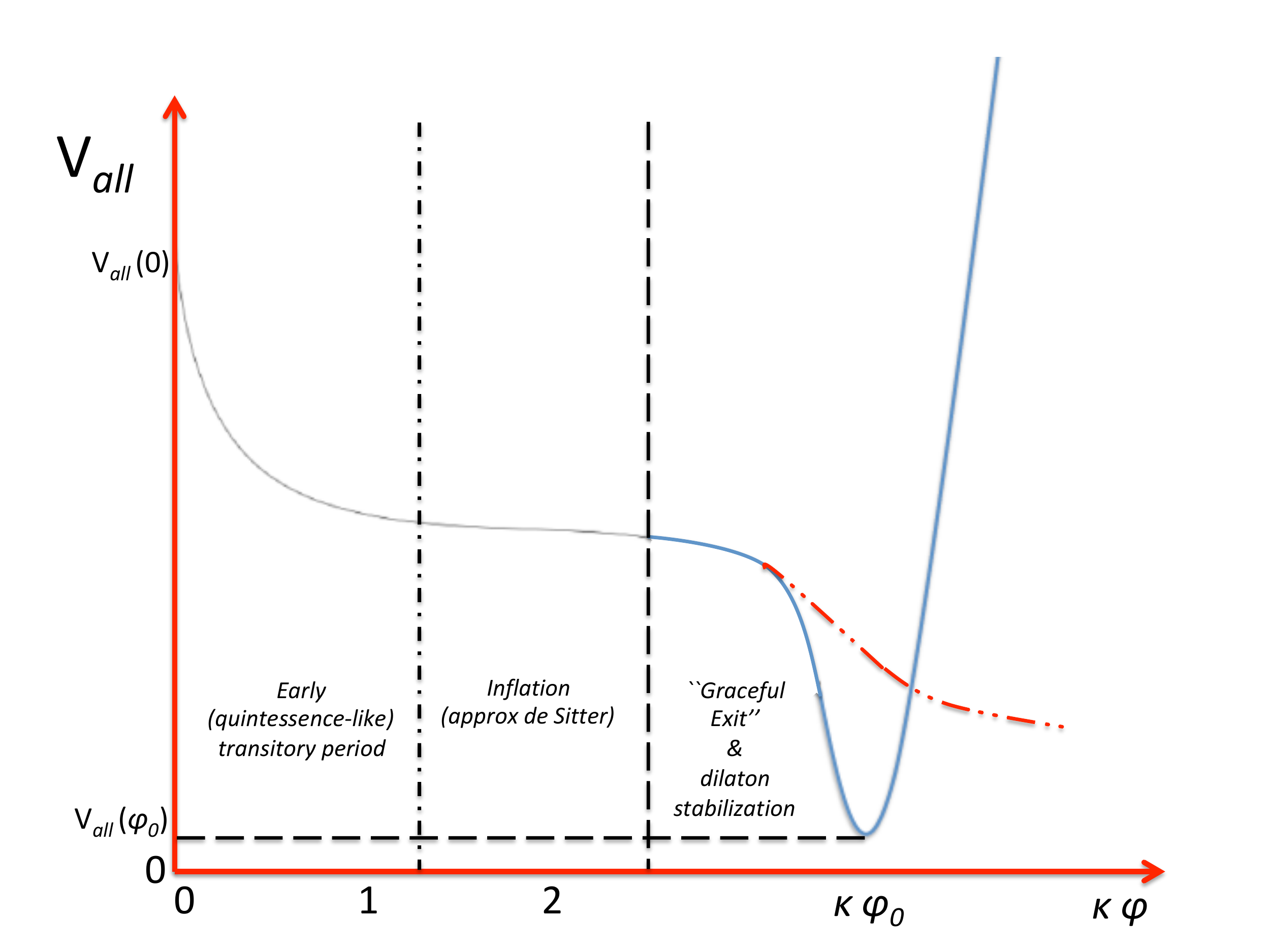}
       \caption{\it A typical non-critical string cosmology potential in the range $0 < \phi < \infty$. The Universe starts with a 
quintessence-like behaviour for small positive dilaton values, and then passes through an approximately  de Sitter era 
for larger values of $\kappa \phi = {\cal O}(1)$, before exiting into an era where the dilaton is stabilised with a value $\phi_0 \gg 1$. 
The exit phase (blue curve, to the right of the dashed vertical line) can be characterised by a dilaton-induced Starobinsky-type potential as 
in the scenario of \cite{ncstaro}. In other non-critical string scenarios~\cite{ncaccel}, the dilaton evolves in such a way that the 
potential exhibits a decaying behaviour, $V_{all} \sim e^{-\tilde c_1 \phi}$ with $\tilde c_1 > 0$, at very large times 
(indicated by the doubly-dotted-dashed red curve), and reaches asymptotically a logarithmic behaviour corresponding to the
linearly-expanding Universe of \cite{aben}, which constitutes an asymptotic ``equilibrium'' situation for non-critical string 
cosmology~\cite{nc,diamant,ncaccel}. }
\label{fig:swamppot}
\end{figure}

In Eq.~(\ref{Gfidef}) above, $q$ denotes the deceleration parameter of the non-critical string Universe $q \equiv - \ddot{a} a /
{\dot{a}}^2$ as function of the time.  We shall be interested in (almost) de Sitter space-times for which
\be\label{qdS}
q \; \simeq \; - 1.
\ee
The potential appearing in (\ref{eqall}) above is defined by 
\be\label{pot}
{{V}}_{all} (\phi) =  \Big(Q^{\;2} (t) \exp{\;( -\sqrt{2} \, |c_0|\, \phi )}+V_0 \Big),
\ee
 where $c_0$ is a constant to be determined by consistency with the modified swampland criteria, and   
for generality we have allowed  for a potential term in the string action $- \sqrt{-G}\;V_0$ in addition 
to that dependent on the central charge deficit term. We expect that $V_0$ might be generated by, e.g., string-loop effects, or else
by interaction with D-particle defects~\cite{ncstaro}, which themselves provide a source of non-criticality. 
In our convention, inflation is obtained for large positive values of the field $\phi$, which correspond to early cosmic times. 

The form of the potential \eqref{pot} is indicated in Fig.~\ref{fig:swamppot}. The $Q$-dependent part of the potential is typical of 
non-critical string cosmologies, and is associated with the central-charge surplus of the underlying world-sheet Liouville theory~\cite{nc,aben}. 
Indeed, for constant $Q$ (as, e.g., in the case of \cite{aben}) and $V_0=0$ the potential exhibits a standard quintessence-like behaviour, 
which is compatible with the First Swampland Conjecture (FSC) \eqref{SC1a}, provided 
$c_0$ is of order one. For instance, the case of a {linearly-expanding} universe with four uncompactified target-space dimensions
studied in \cite{aben} has $c_0=1$, and thus falls in this category. 

The problems with the FSC start when one considers a large-field de Sitter phase (no matter how brief) within the EFT framework, 
for which the dilaton $\phi > 1/\kappa$, $V > 0$ but also, as we shall discuss~\cite{diamant}, $Q(t)$ is no longer constant. 
In fact, the purpose of this paper is to discuss the swampland criteria for non-critical string potentials 
of the form depicted in Fig.~\ref{fig:swamppot}. 
Depending on the details of the $V_0$ part of the potential, the Universe may enter a 
(relatively brief) de Sitter era, before exiting into either (i) an era in which the dilaton is stabilised at a constant value $\phi_0$, 
as in dilaton-induced Starobinsky-like inflation in the context of the non-critical string/brane scenario of \cite{ncstaro} 
(with the value of the potential $V_{all}(\phi_0)$ at its minimum being either zero, or small and positive so as to 
match the current cosmological constant value), or (ii) an era~\cite{ncaccel} in which the dilaton potential exhibits 
another non-critical string quintessence form, $V_{all} \sim e^{-\tilde c_1 \kappa \phi}$, with $\tilde c_1 > 0$ some
positive constant of ${\cal O}(1)$, and the dilaton
evolves with the cosmic time as
\be\label{philna}
\phi \propto \ln a \, , 
\ee
where $a$ is the scale factor of the Universe. This solution is asymptotic for very large cosmic times to the solution of \cite{aben}, 
in which the dilaton scales logarithmically with the Einstein-frame cosmic time, and the Universe is linearly expanding in this frame.~\footnote{It 
should be noted that, if the expansion is measured in string rods, the linear expansion is not really an expansion of the Universe, 
but being an asymptotic equilibrium point in the non-critical string evolution~\cite{ncaccel}, this is not an issue.}.

The dilaton energy density and pressure are 
\bear \label{dileos}
&& \varrho_{\phi}\;=\;\frac{1}{2}\;
{\dot{\phi}}^2 +{{V}}_{all}(\phi)  \, . \nonumber \\
&&p_{\phi}\;=\;\frac{1}{2}\; {\dot{\phi}}^2 -{{V}}_{all}(\phi)  \, .
\eear
We observe in \eqref{dileos} that in a de Sitter background of the type we are interested in, 
under the assumption that the potential dominates, one can ignore the $\dot \phi^2$ terms in front of $V_{all}$, 
so the field $\phi$ acts as a quintessence field inducing an (almost) de Sitter space-time, with
\be\label{dSeos}
\rho_\phi \simeq - p_\phi \simeq V_{all}(\phi) \, . 
\ee
The first of Eqs. \eqref{eqall} is actually the analogue of the Friedmann equation for non-critical string cosmologies. 
Using \eqref{Gfidef} and $\tilde\varrho_m=\tilde p_m=0$,
we may write it in the following form in the de Sitter phase:
\be\label{ncfried}
 H^2 \simeq \frac{1}{3} \, V_{all}(\phi) - \frac{1}{3} \Big(\frac{1}{2\sqrt{2}}\, \ddot{\phi} - \frac{1}{4} \, 
{\dot{\phi}}^2 + \frac{Q(t)}{2\sqrt{2}}\, e^{-\frac{1}{\sqrt{2}}\, \phi} \, \dot{\phi}\Big) \, .
\ee
During the de Sitter phase, we may make the reasonable approximation that the behaviour \eqref{philna} 
characterises the dilaton field. In particular, we assume the proportionality constant in \eqref{philna} to be that 
in the solution of \cite{aben}, which actually describes the asymptotic exit phase from the non-critical string inflation 
model of \cite{diamant}. In such a solution one has: 
\be\label{phidotH}
\dot \phi \simeq \sqrt{2}\, H \quad \Rightarrow \quad  \ddot \phi \simeq \sqrt{2} \dot H \, ,
\ee
which, on account of the second of the equations \eqref{eqall}, implies 
\be\label{hdot}
\dot H \simeq -\frac{3}{2} \, H^2 \, .
\ee
From this we easily obtain, using \eqref{phidotH}:
\be\label{hhdot}
\frac{\dot H}{H^2} = \sqrt{2} \frac{H^\prime}{H}  \simeq -\frac{3}{2} \, H^2 \quad \Rightarrow \quad H \simeq H_I \, e^{-\frac{3}{2\sqrt{2}}\, \phi}~, 
\ee
where the prime denotes differentiation with respect to $\phi$, and $H_I$ is an inflationary scale to be determined
phenomenologically.  

We now observe that, because of \eqref{phidotH} and \eqref{hdot}, the Curci-Paffuti equation \eqref{CXP} implies that
\be\label{solGf}
\tilde{\cal{G}}_{\phi} = c_4 \, ,
\ee
a constant of mass dimension [+2], during the de Sitter phase, which is realised for large values of $\phi$ (see Fig.~\eqref{fig:swamppot}).
Using \eqref{Gfidef} and \eqref{hhdot}, this yields the following evolution with respect to the cosmic time,
or equivalently $\phi(t)$, of the central charge deficit $Q(t)$~\cite{diamant}:
\be\label{qevol}
Q(t) \, H_I \simeq (-c_4) \, e^{\frac{5}{2\sqrt{2}}\, \phi} + \frac{H_I^2}{2} \, e^{\frac{3}{2\sqrt{2}}\, \phi} \, . 
\ee
We now remark that, although the supercriticality of the string corresponds to $Q^2(t) > 0$, we may without loss of generality
assume that $Q(t) > 0$, which allows us to take 
\be\label{posc1}
c_4 = - |c_4| \, \le \, 0.
\ee
As we see later, this is a self-consistent choice when formulating the modified swampland conditions for supercritical strings.

Then, we obtain from \eqref{ncfried}
\be\label{friednc}
H^2 \simeq \frac{4}{7}\, V_{all}(\phi) -   \frac{6}{7}\, Q(t) \, H \, e^{-\frac{1}{\sqrt{2}}\, \phi} \, .
\ee
Using the Bousso bound \eqref{GH}, or equivalently \eqref{GH2}, which is valid independently of the criticality or otherwise of the string, being a geometric
bound on the entropy of states, one may obtain from \eqref{phidotH} and \eqref{friednc}:
\bear\label{ncbouso}
V_{all}(\phi)  &\le & \frac{7}{4}\, \Big(\frac{8\pi^2}{N^\gamma}\Big)^{2/(2-\delta)} +  \frac{3}{2}\,Q(t)\, H\, e^{-\frac{1}{\sqrt{2}}\, \phi}
\nonumber \\
&\equiv&  \frac{7}{4}\, \Big(\frac{8\pi^2}{N^\gamma}\Big)^{2/(2-\delta)} + \mathcal D_\phi (\phi).
\eear
where
\be\label{calDdef}
\mathcal D_\phi (\phi) = \frac{3}{2}\, |c_4| + \frac{3H_I^2}{4} \, e^{-\frac{1}{\sqrt{2}}\, \phi}\,,
\ee
as can be inferred from \eqref{qevol} and \eqref{posc1}. 
It is clear that the presence of the function $\mathcal D_\phi (\phi)$ which depends on the dilaton may drastically modify  
the swampland bound, depending on the details of the non-critical string model, in particular the values of the parameters $c_0$ and $c_4$.

Indeed, deep in the de Sitter phase, the field $\phi$ takes on trans-Planckian values,
for which the distance conjecture \eqref{dSC1} is violated. 
We may also assume that most of the states that are becoming light behave either in a point-like way, corresponding to $\delta \simeq 0$, 
or are string-like, with $\delta =1$.~\footnote{Surface-like states with $\delta=2$ correspond to a singular limit
 of the $N$-dependent terms on the r.h.s. of \eqref{ncbouso} and are subleading, carrying zero measure in the space of states.} 
 Hence, without loss of generality we may assume $0 < \delta < 2$  as a physically relevant range of this 
 parameter for our non-critical string cosmologies, which is also in agreement with a plethora of critical string models~\cite{SC2a}. 

Taking into account the fact that in this large-dilaton regime
there is a large number of states \eqref{tower}
that are light, since $n(\phi) \gg 1$, with $b\, \gamma ={\mathcal O}(1) > 0$, our analysis indicates
that the term $\mathcal D_\phi(\phi)$, proportional to $Q(t)$, dominates the $N$-dependent entropy term on the r.h.s of the inequality \eqref{ncbouso}.
Therefore, using \eqref{calDdef} we may approximate the r.h.s. of the inequality in the second line of \eqref{ncbouso} by 
\be\label{ncbouso2}
 V_{all}(\phi)  \, \lesssim \,  \frac{3}{2}\, |c_4| + \frac{3H_I^2}{4} \, e^{-\frac{1}{\sqrt{2}}\, \phi}\, ,
\ee 
where we stress once again that the (dimensionful) constant $c_4$ is connected to the supercritical-string  central-charge surplus $Q$,
and vanishes for critical strings: $c_4=0$.

The potential \eqref{pot} may be written as
\be\label{pot2}
V_{all} = c_4^2 \, e^{(5-2|c_0|) \, \frac{1}{\sqrt{2}}\, \phi} + \frac{H_I^4}{4} \, e^{(3-2|c_0|) \, \frac{1}{\sqrt{2}}\, \phi} + 
H_I^2 \, |c_4| e^{(4-2|c_0|) \, \frac{1}{\sqrt{2}}\, \phi} + V_0 \, .
\ee
For consistency with the (quintessence-like) exponential behaviour of the potential depicted 
in Fig.~\ref{fig:swamppot},  which we take as our prototype of supercritical string cosmology, 
it {suffices} to consider potentials \eqref{pot2} for which either {(i)} all of the exponents of the exponentials are negative, or {(ii)} 
the largest of them vanishes, and the other two are negative, i.e., 
\be\label{suffcond}
|c_0| \ge 5/2. \\
\ee

\noindent
$\bullet$ {\it Case (i): All the exponents in the exponential are negative} \\
For large positive $\phi$, the leading term in the potential is that with coefficient  $c_4^2$. 
We then see from \eqref{pot} and \eqref{ncbouso} that for consistency with inflationary phenomenology,
a {sufficient} condition for a de Sitter phase at large $\phi$ is 
\be\label{Vcon}
0 \, < \, V_0  \sim H_I^2 \lesssim   \frac{3}{2}\, |c_4|.
\ee
In this way satisfying the Second Swampland Criterion becomes almost trivial. Indeed, 
taking the logarithms on both sides of \eqref{ncbouso2}, differentiating with respect to the field $\phi$, 
and recalling that in our case $V_{all}^\prime < 0$ during and before the inflationary phase (see Fig.~\eqref{fig:swamppot}), 
we obtain 
\be\label{nceSC1b}
\frac{|V_{all}^\prime |}{V_{all}} \,  \gtrsim  \, 
\frac{3\, H_I^2}{4} \, e^{-\frac{1}{\sqrt{2}}\, \phi}\, \frac{1}{ \frac{3}{2}\, |c_4| + \frac{3H_I^2}{4} \, e^{-\frac{1}{\sqrt{2}}\, \phi}}
+ \mathcal O\Big(n(\phi)^{-\frac{2\gamma}{2-\delta}}\,  e^{-\frac{2b\gamma}{2-\delta}\phi}
\Big)
\,.
\ee
which provides a {\it much weaker} constraint on the potential than \eqref{SC1a} for large $\phi \gg 1$. 
We reiterate that above we assumed the presence of a large number of  light string states, 
in which case the terms $\mathcal O\Big(\dots \Big)$ on the r.h.s. of the inequality \eqref{nceSC1b} 
are subleading for $b, \, \gamma ={\mathcal O}(1) > 0$, as explained previously.
Otherwise, the standard swampland condition is recovered, which is the case for critical strings.

We now check the conditions on the parameters of the potential \eqref{pot2} under which it satisfies the condition \eqref{nceSC1b}
in the large-$\phi$ (de Sitter) regime (see Fig.~\ref{fig:swamppot}). To this end, we compute the quantity:
\be\label{conpot2} 
\frac{|V_{all}^\prime|}{V_{all}} \simeq \frac{|5-2|c_0||}{\sqrt{2}} \, c_4^2 \, e^{\frac{(5-2|c_0|)}{\sqrt{2}}\, \phi} \, 
\frac{1}{V_0 + c_4^2 \, e^{\frac{(5-2|c_0|)}{\sqrt{2}}\, \phi} + \dots} \, ,
\ee
where the $\dots$ denote subleading terms in the large-field de Sitter regime $\phi \gg 1$
with $|c_0| > 5/2$. Comparing \eqref{conpot2} with \eqref{nceSC1b}, we observe that \eqref{nceSC1b} is satisfied 
provided:
\bear\label{c06}
5-2|c_0| &=& -1 \quad \Rightarrow \quad |c_0| = 3, \quad {\rm and} \nonumber \\ 
\frac{c_4^2}{V_0} \, &\gtrsim & \frac{1}{\sqrt{2}} \, \frac{H_I^2}{|c_4|} \, ,
\eear
which, since $V_0 \sim H_I^2$ \eqref{Vcon}, implies: 
\be\label{c4HI}
\kappa^2 \, |c_4| \, \gtrsim \, 0.89 \, \Big(H_I \, \kappa\Big)^{4/3} \, ,
\ee
where we have reinstated the factors of $\kappa$ units, for clarity.
On the other hand, from the Bousso bound on the potential (\eqref{Vcon} or equivalently \eqref{ncbouso2}), we must have
\be\label{bousoc4HI}
\kappa^2 \, |c_4| \, \gtrsim \, 0.67 \, \Big(H_I \, \kappa\Big)^{2}
\ee
for $\kappa \, \phi \gg 1$. Taking into account the fact that phenomenologically  $H_I \, \kappa \sim 10^{-4} \, \ll \, 1$~\cite{planck}, 
we find that a 
 stronger lower bound on $|c_4|$ comes from \eqref{c4HI}, which is the result of the 
modified swampland condition \eqref{nceSC1b} in the supercritical string model. \\

\noindent
$\bullet$ {\it Case (ii): One exponent vanishes} \\
Let us now  consider the case in which one exponent vanishes, and the other two are negative with 
\be\label{cosat}
|c_0| = \frac{5}{2}.
\ee
In this case, we see from \eqref{pot} and \eqref{ncbouso} that for consistency with inflationary phenomenology,
a {sufficient} condition for obtaining a de Sitter phase at large $\phi$ is
\be\label{Vcon2}
0 \, < \, |c_4|^2 + V_0  \sim \kappa^{-2} \, H_I^2 \lesssim   \frac{3}{2}\, |c_4| \kappa^{-2} \, ,
\ee
where we have again reinstated the factors of $\kappa$, for clarity.
Performing a similar analysis as before, i.e., by computing $V_{all}^\prime/V_{all}$ from \eqref{pot2} at large $\phi \gg 1$, 
but now under the condition \eqref{cosat}, 
we reach the conclusion that in order for 
$V_{all}$ to satisfy \eqref{nceSC1b} and \eqref{Vcon2} simultaneously 
one must have
\bear\label{Vcon3}
\frac{|c_4| \, H_I^2}{|c_4|^2 + V_0} \, \gtrsim \, \frac{1}{\sqrt{2}} \, \frac{H_I^2}{|c_4|} \, \quad &\Rightarrow& \quad  |c_4|^2 \, \gtrsim \, \frac{|c_4|^2 + V_0}{\sqrt{2}} \sim \frac{H_I^2\, \kappa^{-2}}{\sqrt{2}}, \nonumber \\
&{\rm and}& \qquad  |c_4|^2 \, \gtrsim \frac{4}{9} \, H_I^4\,. 
\eear
The stronger bound on $|c_4|$ is again that coming from the modified swampland constraint \eqref{nceSC1b} on the  slope of the dilaton potential.\\

{\it The condition \eqref{nceSC1b} is necessary and sufficient for the validity of the EFT during the de Sitter phase of
supercritical string cosmology. The latter contains a plethora of stringy states, which 
result in non-criticality, but collectively their presence is taken into account by the presence of the EFT terms  on 
the r.h.s. of \eqref{eqall}, stemming from the identification of the Liouville mode with the cosmic time~\cite{nc}.
The modified criterion \eqref{ncbouso2} or equivalently \eqref{nceSC1b} is compatible with slow-roll inflation, 
in contrast to the First Swampland Conjecture \eqref{SC1a}. The latter is recovered in the critical string limit $Q \to 0$, 
since the $N$-dependent term on the r.h.s. of the inequality in \eqref{ncbouso} is the only term present in that case.}

\vspace{0.2cm}
\subsection*{Embedding in microscopic conformal field theory models} 

\nicktext{The constants
$c_4$ and $c_0$ entering the effective potential \eqref{pot2} would be determined by the embedding of the effective supercritical string cosmology model
into a microscopic string realisation. For instance, one may consider the framework of \cite{aben}, in which the internal dimensions correspond to minimal models
characterised by specific (discrete) values of their central charges. In such a case, $c_0$  and $c_4$ are fixed by the requirement that 
the supercritical string model approaches the linearly-expanding Universe model of \cite{aben} asymptotically as $\phi \to \infty$. 
In this connection, we observe that setting $V_0=0$ and 
\be\label{c0det}
 |c_0| = 7/2 
 \ee
 in \eqref{pot2} (which is compatible with \eqref{suffcond}),  the potential assumes the form 
\be\label{pot3}
V_{all} = c_4^2 \, e^{-\sqrt{2}\, \phi} + \frac{H_I^4}{4} \, e^{-\frac{4}{\sqrt{2}}\, \phi} + 
H_I^2 \, |c_4| e^{-\frac{3}{\sqrt{2}}\, \phi}  \quad \stackrel{\phi \to \infty}{\Rightarrow} \quad  V_{all}(\phi \to \infty) \simeq \, c_4^2 \, e^{-\sqrt{2}\, \phi} \, . 
\ee
The asymptotic form of $V_{all}$ coincides with the dilaton potential in the model of \cite{aben}, 
when we identify $c_4^2$ with the central-charge surplus $\delta c \propto Q_\star^2$= constant. }

\nicktext{We recall that, in the approach of \cite{aben}, if $c_{\rm tot}$ denotes the total central charge of the world-sheet $\sigma$-model theory, and $d$ the number of uncompactified string dimensions, one has~\cite{aben} 
\be\label{ctot}
c_{\rm tot} = d  - 12\, Q_\star^2 + c_I \,, 
\ee
where $c_I$ is the central charge corresponding to the conformal field theory of the internal manifold. In the bosonic prototype string theory, decoupling of reparametrisation ghosts requires $c_{\rm tot}$=26 (10 for superstrings), and in our case $d=4$. This implies that 
\be\label{scdef}
12\,Q_\star^2= c_I - 22 \,, 
\ee
which is a model-dependent constant in the model of \cite{aben} that depends on the conformal data of the internal theory, specifically on its central charge $c_I$.  
Supercritical strings correspond to $Q_\star^2 > 0$.~\footnote{The terminology supercritical refers to the fact that in terms of a string in $D$ flat uncompactified target space time dimensions, $D=d + c_I > 26 $ (or 10 for superstrings).} In terms of the target-space effective action, in the model of 
\cite{aben} there is a positive vacuum energy of quintessence type of the form
\be\label{qeff}
\int d^4x \sqrt{-g} \, 4 Q_\star^2 e^{-\sqrt{2}\phi} 
\ee
in our conventions \eqref{standard}. 
Comparing \eqref{pot3} with \eqref{qeff}, we observe 
that the value of $c_4$ is
fixed by the choice of conformal field theory corresponding to the internal manifold degrees of freedom: 
\be\label{pot4}
c_4^2 = 4\, Q_\star^2 ,
\ee
where $Q_\star^2$ is the central charge surplus,~\footnote{It corresponds to a Wick-rotated Coulomb charge at infinity for the minimal conformal field theory models considered in \cite{aben} as a description of the internal-dimension manifolds of the super-critical string.} which is a rational real number in the models of \cite{aben}. }

\nicktext{ Setting $\phi = 0$ in \eqref{qevol} and taking \eqref{posc1} into account, we find
\be\label{qevol0}
Q(\phi=0) = \frac{|c_4|}{H_I} + \frac{1}{2} \, H_I~, 
\ee
where $H_I $ is expressed in units of $\kappa^{-1}$. For realistic inflationary scenarios~\cite{planck}
$H_I \kappa \lesssim 10^{-4} \ll 1$, so  in Planck units $H_I \ll 1$ in \eqref{qevol0}, \eqref{qevol}, 
and hence $Q^2(0) > Q^2_\star = Q^2(\phi \to +\infty)$ when \eqref{pot4} is taken into account. One may consider
\eqref{qevol0} as corresponding to an  initial fixed-point Conformal Field Theory(CFT), e.g., another minimal model, following \cite{aben}, with 
\be\label{inter}
Q^2_{\star \rm 1~CFT} \equiv Q^2(\phi = 0) > Q_\star^2 (\phi \to \infty) \equiv Q^2_{\star \rm 2~CFT} =\frac{|c_4|}{4}\,.
\ee 
In the minimal models considered in \cite{aben} as providing conformal models for the internal
 manifold, $Q^2_{\star \rm i~CFT}$, $i=1,2$ take  {discrete} rational values. }

\nicktext{ Thus $Q^2(t)$ \eqref{qevol},
which defines the overall ``running'' central charge surplus in the supercritical string model including the contributions from the internal manifold, {interpolates} \eqref{inter} between specific minimal models of the type considered in \cite{aben}.  We note that 
the evolution \eqref{qevol} of $Q^2(t)$ is independent of the constant $c_0$, but assumes implicitly that there exists a slowly-running inflationary phase at a scale $H_I$. This is not the case in the model of \cite{aben}, which describes a linearly-expanding universe, but it is consistent with the aforementioned r\^ole of this theory as an asymptotic fixed point of the supercritical string model~\cite{timeRG}. }

\nicktext{ In this approach one considers an interpolating theory with a central-charge surplus  $Q^2(t)$ dependent on the Liouville mode that is identified with the cosmic time~\cite{timeRG}, hence there is no overall increase in the target-space dimensionality of the interpolating string model. There is a departure from criticality for the two-dimensional (world-sheet) field theory corresponding to the internal manifold of the string. During the interpolating phase, the internal conformal field theory  at $\phi=0$, corresponding to an initial cosmic time, is perturbed by relevant operators in a world-sheet renormalisation-group sense. This induces a flow between the two fixed points described by  the conformal field theories corresponding to the Coulomb charges $Q^2_{\star \rm i ~CFT}$, $i=1,2$  \eqref{inter}. }
 
\nicktext{ During the flow, Liouville dressing is required  to maintain $c_{\rm tot}=26$, 
as required for ghost decoupling.
In this way,  the internal central charge $c_I \rightarrow c_I(t)$ becomes itself a function of the Liouville mode (i.e., cosmic time), so that \eqref{scdef} (or, equivalently, \eqref{ctot}) is satisfied with $c_{\rm tot}$=26.   The theory flows asymptotically to a fixed-point theory corresponding to a dilaton potential of the form \eqref{pot3} for large positive values of $\phi \to \infty$, where the model asymptotes to one of the conformal field theory models of \cite{aben}.  In this scenario the constants $c_4^2$ \eqref{pot4} and $c_0$ \eqref{c0det} are determined. }

\nicktext{ The above serves as an example of how one can determine the relevant constants appearing in the effective dilaton potentials \eqref{pot} of the supercritical string cosmology models by means of specifying the underlying microscopic world-sheet conformal field theory models. 
As we have emphasized, the models of \cite{aben} do not describe, as they stand, inflationary physics. They serve as asymptotic fixed points in a world-sheet renormalisation-group sense of a more complete theory that describes inflation for relatively small values of the dilaton $\phi > 0$, as happens in the model \eqref{pot2} examined above.   
More complicated super-critical string models also exist in the literature, such as the type-0 string non-critical string theory model of \cite{papant} and the tachyon-dilaton models of \cite{kostouki}. Such models are characterised by transient inflationary phases, during which the central charge surplus diminishes as the cosmic time increases as in our case \eqref{pot2} above, before oscillating around zero
for a short period of time, and then asymptoting to one of the minimal model values of the linearly-expanding Universe of \cite{aben}. }

\nicktext{ We stress again that our considerations deal with generic supercritical effective theories,
 without an emphasis on presenting detailed microscopic string constructions. Our approach should be viewed  as an attempt to demonstrate
 that generically, supercritical string with a Liouville-renormalized central charge surplus $Q^2(t)$ can lead to cosmologies that are embeddable 
 in a UV-complete theory of quantum gravity, in which the corresponding EFT bypasses
 the swampland criteria. The aforementioned specific string models serve as concrete examples of microscopic string theory models where this situation is realised explicitly. }
 
\nicktext{ Since our supercritical string (Liouville) cosmology  is viewed as an {interpolation}  between critical string models, during the monotonic Liouville evolution taken to correspond in our convention to a dilaton field that increases with the cosmic time, the descent from the UV of massive string states that become massless during this evolution, and thus fail to decouple, is governed by the same criteria as in the critical string cosmologies used in \cite{SC2a} and above when formulating the second swampland conjecture. In particular, the generic parametrisation \eqref{entropy} of the entropy of massive string states is valid in our 
 Liouville non-critical string case as well, with the Hubble parameter depending on the time-like Liouville mode that is identified with the cosmic time in our approach~\cite{timeRG}. }
 
\nicktext{ This dependence parametrizes {collectively} the failure of decoupling of the massive states in the following sense, familiar from our prototype example of a non-critical string theory, realised in the two-dimensional stringy black hole system~\cite{emn}:  (i) one starts from an underlying conformal field theory, corresponding to an initial value of the dilaton field; (ii) then, some specific perturbations, depending on the microscopic content of the theory, move it away from criticality. The perturbations may mix massless and massive states in the spectrum,  such that some of the massive stringy states fail to decouple as the dilaton evolves, exactly as happens in the aforementioned black hole example~\cite{emn}.
This failure is encoded in the non-criticality of the string cosmology model that interpolates between critical string backgrounds, and it is in this sense that the effective description of the cosmology in terms of local field theories at low energies, including the dilaton and graviton fields, suffices. Any non-locality stemming from these massive states is described collectively by the effects of Liouville mode. The embedding of such an EFT into a consistent quantum gravity model is guaranteed by bypassing the swampland criteria, as a consequence of the modified cosmological evolution that characterises the supercritical string cosmologies~\cite{diamant}. }

\nicktext{ This is the essence of our supercritical string cosmology~\cite{emn,timeRG} and its evasion of the swampland criteria, as presented above. }

\subsection{The Special Case of a Dilaton-Induced Starobinsky-Like Potential~\cite{ncstaro}}

In light of the above discussion, we infer that
the dilaton/D-particle-induced Starobinsky-like potential of \cite{ncstaro} constitutes a consistent non-critical string cosmology,
as it also satisfies \eqref{nceSC1b}. 
However, for completeness we should mention that this model
is constructed somewhat differently, in that
its dilaton central charge deficit is assumed to be subcritical, as the brane world lives in $d=4$ dimensions, 
and the non-criticality arises from the interaction with ensembles of D-particles living on the brane world, 
which enter the latter from the bulk as the brane Universe propagates in cosmic time. 

The corresponding effective action in the Einstein-frame is given by~\cite{ncstaro} 
\be\label{nceastaro}
S_{\rm eff}^{\rm staro}  = \frac{1}{2\kappa^2} \, \int d^4x \sqrt{-g} \Big(  R(g) - (\partial_\mu \phi)^2 - V_{\rm Nc-Staro} 
+ \dots \Big) \, ,
\ee
where the four-dimensional gravitational constant is given by $\kappa^2 = 8\pi \alpha^\prime \, \mathcal V_C^{-1}$, 
and $\alpha^\prime = M_s^{-2}$ is the Regge slope where $M_s$ is the string mass scale, 
and $\mathcal V_C$ is the volume of the extra compact dimensions (or the volume of the portion of the bulk space between 
the brane worlds, depending on the construction). 
The potential  $V_{\rm NC-Staro}$ is given by~\cite{ncstaro}
\be\label{staro}
V_{\rm NC-Staro} = \mathcal A - \frac{44}{3\kappa^2 \, \alpha^\prime} e^{-\sqrt{2}\kappa \, \phi} + \dots  \ge 0 \, . 
\ee
The quantity 
 $\mathcal A  = 16\pi \frac{n\, \sqrt{\alpha}}{g_s^{(0)} \, \mathcal V_c}  > 0$, with $g_s^{(0)} < 1$,  
 depends on the (three-)density of D-particles on the brane world $n$, which is assumed to be approximately constant, 
with its dilution due to cosmic expansion being compensated by an influx of D-particles from the bulk. 
The density $n$ and masses of the D-particles can be adjusted so that the potential vanishes at $\phi=0$. 
The magnitude of the quantity $A$ provides the inflationary scale in the model. 

We now make some important remarks concerning the structure of such a potential. 
Notice that in the conventions of \cite{ncstaro}, the potential {increases} for increasing {positive} $\phi$, which
 is a crucial difference from the non-critical string potentials \eqref{pot2}. 
On the other hand, as in the supercritical-string case, the inflationary phase occurs for large positive $\kappa \, \phi \gg 1$. 
Nonetheless, because now $V_{\rm NC-Staro}^\prime > 0$, some crucial modifications need to be made in the non-critical string analysis 
presented above that led to the swampland conditions. 

The potential \eqref{staro} corresponds to \eqref{pot}, but now $Q^2 < 0$ as in sub-critical strings. Formally, 
the non-critical string analysis can be performed in such a case by analytically continuing to Euclidean cosmic time 
\be\label{ancon}
t \to i t\,, \quad Q \to i Q\, , \quad H \to i H \,, 
\ee
maintaining otherwise the structure of \eqref{ncfried}. 
Using the analytic continuations of the expressions \eqref{hdot} and \eqref{hhdot} and the 
Curci-Paffuti equation we arrive at a modified evolution equation for the central charge deficit $Q(t)$, which 
results in a modified Bousso bound when we revert to Minkowski cosmic time:
\be\label{ncbouso2staro}
 0 < V_{\rm NC-Staro}(\phi)  \, \lesssim \,  \frac{3}{2}\, |c_4| - \frac{3H_I^2}{4} \, e^{-\frac{1}{\sqrt{2}}\, \phi}\, ,
\ee 
where we maintain the same sign in front of the constant $c_4$ as in the supercritical string case \eqref{ncbouso2}, as required  
by the positivity of the potential (the reader can check from \eqref{solGf} that we have the freedom to do so).
However the sign of the $H_I^2$-dependent term is now the opposite, in view of \eqref{ancon}. 

Taking the logarithm of \eqref{ncbouso2staro} and differentiating with respect to $\phi$, we now have 
\be\label{SC1Staro}
0 \, <  \, \frac{V_{\rm NC-Staro}^\prime }{V_{\rm NC-Staro}} \,  \lesssim  \, 
\frac{3\, H_I^2}{4} \, e^{-\frac{1}{\sqrt{2}}\, \phi}\, \frac{1}{ \frac{3}{2}\, |c_4| - \frac{3H_I^2}{4} \, e^{-\frac{1}{\sqrt{2}}\, \phi}}
+ \mathcal O\Big(n(\phi)^{-\frac{2\gamma}{2-\delta}}\,  e^{-\frac{2b\gamma}{2-\delta}\phi}
\Big)
\,.
\ee
From this we see that the condition \eqref{nceSC1b} is now modified by changing the sign of the inequality, since 
now the slope of the (positive) potential is {also} positive $V_{\rm NC-Staro}^\prime > 0$. 

It is thus straightforward to see that, during the trans-Planckian inflationary de Sitter phase with $\kappa \, \phi \gg 1$, 
the Starobinsky-like potential \eqref{staro} satisfies the modified Swampland conjecture \eqref{SC1Staro} quite comfortably. 
Indeed, we see that
\be\label{staroder}
\frac{V^\prime_{\rm NC-Staro}}{V_{\rm NC-Staro}} = \sqrt{2} \frac{44}{3 \kappa^2 \, \alpha^\prime} \, \frac{e^{-\sqrt{2}\kappa\, \phi} }{\mathcal A - \frac{44}{3\kappa^2 \, \alpha^\prime} e^{-\sqrt{2}\kappa \, \phi} + \dots } \, ,
\ee{
which can be compared with the r.h.s. of the inequality \eqref{SC1Staro}. The r.h.s. of \eqref{staroder} is much smaller
for phenomenologically relevant values of  $H_I, \mathcal A, \alpha^\prime, |c_4| ={\mathcal O}(A)$,
in  view of the difference in the powers in the exponents of the exponentials  containing the large $\phi$ field. 
Thus, the condition \eqref{SC1Staro} is {naturally}
satisfied and an EFT description is valid. 

We stress once more that, in our considerations in this work, and more generally in our non-critical 
cosmological string models~\cite{nc,diamant,ncaccel}, the string coupling remains weak: $g_s < 1$
during the entire cosmic evolution  from the origin of cosmic time, which corresponds to $\phi=0$, until the present era, including the 
inflationary epoch (see Fig.~\ref{fig:swamppot}).
Hence the D-particles remain heavy in the scenario of 
\cite{ncstaro}, since their masses are proportional to  
\be\label{Dmass}
M_D \sim g_s^{-1} \, (\alpha^\prime)^{-1/2}\,. 
\ee
and $g_s =g_s^{(0)} \, e^{-\frac{2}{\sqrt{2}} \, \phi} \, < \, 1$. 

Nonetheless, as the dilaton takes on large values, other string states  (e.g., scalar moduli or
stringy extended objects) can descend from the UV to contaminate the 
EFT in the way described above. However, in the context of our supercritical strings this is taken into account by representing 
collectively the ``environment" of such states via extra contributions to the EFT cosmological equations, 
incorporated via the appropriate Liouville dressing and identification of the (time-like) Liouville mode with the cosmic time 
(see the r.h.s. of \eqref{eqall} in the concrete example discussed in the current work).
Thus, we always retain an EFT description, which is reflected in the triviality of the modified 
swampland condition \eqref{nceSC1b} in supercritical string cosmologies. There are, however,
some restrictions on the parameters of the respective dilaton potentials, see, e.g.,
 \eqref{suffcond} and \eqref{Vcon}, needed to match the relative phenomenology.

\section{Discussion: Issues with the Weak Gravity Conjecture \label{sec:WGC}}

Before closing, we discuss the consistency of our results with the 
{Weak Gravity Conjecture} (WGC)~\cite{WGC} and its extension to include scalar fields~\cite{palti}.
The WGC states essentially that gravity is the weakest of the fundamental forces in four dimensions.
According to the WGC, if a theory contains a U(1) gauge symmetry with coupling $g$, then there must exist a state of charge $q$ and mass $m$, such that~\cite{WGC} 
\be\label{WGC}
g\, q M_P = m
\ee
where $M_P = \sqrt{8\pi}\, \MPl$ is the four dimensional Planck scale. 
Its scalar field extension~\cite{palti} implies that if $m(\phi_i)$ is the mass gap of some scalar field $\varphi$ (not the dilaton)
interacting with the light scalar moduli fields $\phi_i$ (which include the dilaton) within some four space-time dimensional EFT, 
then gravity remains the weakest force, which requires that interactions involving exchanges of the light scalars
be such that the mass of the scalar field 
$\varphi$ satisfies
\be\label{massWGC}
m^2 \le \tilde g^{ij} \, \partial_i m \, \partial_j m \, ,
\ee
where $\tilde g_{ij}$ is the metric in the space of the moduli fields $\phi_i$. 
As we do not deal explicitly with charged matter in this work,
it is this scalar extension \eqref{massWGC} that is of interest to us here. 

First we notice that in models involving (scalar) D-particles, as in the Starobinsky inflation model of \cite{ncstaro}, 
the masses of the D-particle states themselves satisfy \eqref{massWGC}, since that the D-particle mass is inversely proportional to the 
string coupling \eqref{Dmass}. In terms of the standard dilaton $\Phi$ \eqref{standard}
that  is used in the canonical expression for the 
D-particle mass, the string coupling is given by $g_s =e^\Phi$, so 
we see that condition \eqref{massWGC} is satisfied, in fact it is saturated with a constant `metric' in theory space
that is equal to the identity. However, having said that, the conjecture is formulated at present for local field theory scalars, 
and its extension  to stringy/brany objects is not yet formulated to a level of precision and understanding that allows us to  
apply the above result \eqref{massWGC} with rigour. 

Moreover, in our non-critical string cosmologies, the considerations around the scalar WGC pertain to 
masses of scalars other than the dilaton, which is one of the string moduli. As such, it
is not relevant for the non-critical string dilaton potentials \eqref{pot2}  discussed in this work. 
One needs a formulation of the scalar WGC that incorporates the potential of the (self-interacting) dilaton.~\footnote{In
passing, we mention that the scalar WGC  (\ref{massWGC})
appears to be inconsistent with the postulated properties of axion potentials.}

In this connection, an {improved version} of the scalar WGC has been suggested in 
\cite{ibanez}:
\be\label{improvedWGC}
(V^{\prime\prime})^2  \le \Big( 2  (V^{\prime\prime\prime})^2 - V^{\prime\prime}\, V^{\prime\prime\prime\prime} \Big) \, 
\kappa^{-2}\,,
\ee
which is expressed in terms of the scalar potential $V$, for canonically-normalised fields.
This improved WGC resolves the axion puzzle, leading to a plausible constraint on the axion decay constant, $f_a \le M_P$, as expected.
In  principle, \eqref{improvedWGC} should be applicable to our case, where we deal with a canonically-normalised dilaton 
$\phi $ \eqref{standard}.  

As discussed in Section~\ref{sec:swamp}, in both the cases \eqref{cosat} and \eqref{c06} that are compatible with the 
non-critical string modification of the First Swampland Conjecure \eqref{nceSC1b},  
we are dealing with large dilaton field values $\kappa \, \phi \gg 1 $ and exponential potentials in a de Sitter phase of the Universe:
\be\label{gpot}
V \sim A \, e^{\gamma \, \kappa \, \phi} \, ,
\ee
where $\gamma$ is a negative dimensionless constant that takes the value
\be\label{gdil}
\gamma = -\frac{1}{\sqrt{2}}\,. 
\ee
in both cases.
For such potentials, the condition \eqref{improvedWGC} is 
{\it not}  satisfied.

We remark, though, that there is no rigorous diagrammatic derivation of the improved scalar WGC \eqref{improvedWGC}.
In particular, the coefficients of the various scalar self-interaction terms appearing on the r.h.s. of the inequality 
could in principle be modified. For this reason,  a generalised form of the scalar potential WGC \eqref{improvedWGC} 
has been proposed in \cite{kusenko}:
\be\label{improvedWGC2}
(V^{\prime\prime})^2  \le \Big( 2c^\prime_1  (V^{\prime\prime\prime})^2 - c^\prime_2 V^{\prime\prime}\, V^{\prime\prime\prime\prime} \Big) \, 
\kappa^{-2}\,,
\ee
where $c^\prime_1$ and $c^\prime_2 > 0$ are positive coefficients of order $\mathcal O(1)$.
With appropriate values of these coefficients, the generalised form of WGC \eqref{improvedWGC2} 
can accommodate various scalar potentials of interest in the early Universe, as in Q-ball models, 
but also exponential scalar potentials of the form \eqref{gpot} for $\gamma \le 1$, 
such as our supercitical string case with a canonically-normalised dilaton \eqref{gdil}.
The condition would be saturated in that case for $c_1^\prime = c_2^\prime =2$.

On the other hand, we note that the condition is {\it satisfied}, indeed {\it saturated}, if 
one uses the {standard dilaton} $\Phi$ \eqref{standard} appearing in the definition of the string coupling, 
$g_s=e^{\Phi}$, as the string loop-counting parameter. However, we note that in such a
formalism the dilaton does not have canonically-normalised kinetic terms, but appears in the 
string-inspired low-energy gravitational action 
in a D-dimensional uncompactified target space in the form:
\be\label{ncea}
S_{\rm eff}  = \frac{1}{2\kappa^2} \, \int d^Dx \sqrt{-g} \Big(  R(g) - \frac{4}{D-2} (\partial_\mu \Phi)^2 - e^{\frac{4\, \Phi}{D-2}} \, Q^2  + \dots \Big) \, ,
\ee
where $Q^2$ is the central charge surplus of the supercritical string~\cite{aben,nc}, 
and the $\dots$ denote other fields, higher derivative terms, {etc.}. Thus, for four large uncompactified dimensions, $D=4$,
one obtains the result \eqref{standard} for the canonically-normalised dilaton, $\phi$, used in \cite{ncstaro} and above.

The satisfaction of the improved WGC \eqref{improvedWGC} by $\Phi$ but not $\phi$ is curious, but may be the 
correct form to be used in the case of non-critical strings with exponential scalar potentials. 
We remark that in \cite{ibanez} the conjecture \eqref{improvedWGC} was derived by looking at four-point scalar interactions, 
one with the scalar field as a mediator and the second a four-scalar contact interaction, which could be due to the exchange
of a massive mediator. The four-point interaction is repulsive, and is associated with the term in the conjecture
involving the fourth derivative of the field. Thus, subtracting the contact interaction from the the four-point amplitude 
with a scalar mediator (which is attractive, and corresponds to the term in the conjecture involving the square of the 
cubic derivatives of the potential with respect to the scalar field) should give a result at least as weak as the 
fourth-order interaction mediated by a four-dimensional graviton exchange, which is essentially the square 
of the second derivative of the potential with respect to the field (which defines the mass-squared of the scalar field). 
This is the philosophy behind~\eqref{improvedWGC}. For non-critical strings with exponential potentials the situation 
needs to be revisited, which might lead to an explanation on the aforementioned discrepancy, 
and why the (non-canonically-normalised) form $\Phi$ of the dilaton should be used in the string effective action.

Finally, we note that, for the dilaton-induced Starobinsky potential \eqref{staro},  
in the inflationary phase occurring for large positive $\phi$, and with a canonical kinetic term as in \eqref{standard},  
the condition \eqref{improvedWGC} is trivially satisfied, as can be seen by direct computation.
The exponent $\sqrt{2}$ is crucial for this result. In this model the potential 
\eqref{staro} is of the following generic form for large $\kappa \, \phi \gg 1$: 
\be\label{generic}
V = |c_5| - |c_6|  \, e^{-\sqrt{2}\, \kappa\, \phi} + \dots
\ee 
with $c_5, c_6 \in \mathbb R$ real constants, and with the kinetic term of $\phi$ canonically normalised.
We see from \eqref{generic} that the improved WGC \eqref{improvedWGC} requires $4 \, \kappa^6 \, < 8 \, \kappa^6$,
which is satisfied. We also remark that the condition is satisfied for this model also if one uses the non-canonically
normalised standard dilaton $\Phi$  appearing in the string effective action \eqref{ncea} or \eqref{nceastaro}. 

However, we caution that there are many ambiguities surrounding the formulation of the WGC.
For example, there are other forms of the WGC~\cite{karim} that take into account the weakness of gravitational interactions, 
implying that the situation involving scalars and the WGC is still ambiguous, reflecting once more our ignorance of the structure of quantum gravity. 
Thus, although the conjecture that gravity must be the weakest force 
certainly sounds plausible for a quantum theory of gravity, the rigorous and unambiguous
quantification of this conjecture is nevertheless not trivial.  In particular, it is quite possible that the formulation of the scalar WGC 
\eqref{massWGC} may need modification in the case of a non-critical string cosmology with a running dilaton,
while maintaining the nature of gravity as the weakest observable force in 
our four-dimensional world.  We conclude that there is no dramatic contradiction between the superstring cosmology scenarios we advocate 
and the Weak Gravity Conjecture. Moreover, it seems clear from the discussion in Section III that these models are quite
compatible with the Swampland Conjectures.\\

{\it Affaire \`a suivre/to be continued  \dots}

\section*{Acknowledgements}

The work  of JE and NEM is supported in part by the UK Science and Technology Facilities  research Council (STFC) under the research grant
ST/P000258/1, and JE also receives support from an Estonian Research Council Mobilitas Pluss grant. The work of DVN is supported in part 
by the DOE grant DE-FG02-13ER42020 at Texas A\&M University and in part by the Alexander S. Onassis Public Benefit Foundation. 
JE and NEM participate in the COST Association Action CA18108 ``{\it Quantum Gravity Phenomenology in the 
Multimessenger Approach (QG-MM)}''. NEM also acknowledges a scientific associateship (``\emph{Doctor Vinculado}'') at IFIC-CSIC, 
Valencia University, Valencia, Spain.

\end{document}